\documentclass[tightenlines,a4paper,altaffilletter,twoside,secnumarabic,
               twocolumn,floatfix,nofootinbib]{revtex4}

\usepackage[english]{babel}
\usepackage{amsmath,amssymb}

\allowdisplaybreaks[1]

\newenvironment{refentry}[1]
  {\hangindent=\parindent \hangafter=1 \noindent \texttt{#1} \smallskip}
  {\bigskip}

\newcommand{\diag}{\mathrm{diag}}
\renewcommand{\vec}[1]{\mathbf{#1}}
\newcommand{\mat}[1]{\mathbf{#1}}
\renewcommand{\Re}{\mathrm{Re}}

\begin{document}

\title{Efficient numerical diagonalization of hermitian $3 \times 3$ matrices}
\author{Joachim Kopp}    \email[Email: ]{jkopp@mpi-hd.mpg.de}
\affiliation{Max--Planck--Institut f\"ur Kernphysik, \\
             Postfach 10 39 80, 69029 Heidelberg, Germany}

\begin{abstract}
  A very common problem in science is the numerical diagonalization of symmetric or
  hermitian $3 \times 3$ matrices. Since standard ``black box'' packages may be
  too inefficient if the number of matrices is large, we study several alternatives.
  We consider optimized implementations of the Jacobi, QL, and Cuppen algorithms
  and compare them with an analytical method relying on Cardano's formula for the
  eigenvalues and on vector cross products for the eigenvectors. Jacobi is the most
  accurate, but also the slowest method, while QL and Cuppen are good general purpose
  algorithms. The analytical algorithm outperforms the others by more than a factor
  of 2, but becomes inaccurate or may even fail completely if the matrix entries
  differ greatly in magnitude. This can mostly be circumvented by using a hybrid
  method, which falls back to QL if conditions are such that the analytical
  calculation might become too inaccurate. For all algorithms, we give an
  overview of the underlying mathematical ideas, and present detailed benchmark
  results. C and Fortran implementations of our code are available
  for download from \texttt{http://www.mpi-hd.mpg.de/$\sim$globes/3x3/}.
\end{abstract}

\maketitle

\section{Introduction}

In many scientific problems, the numerical diagonalization of a large number of
symmetric or hermitian $3 \times 3$ matrices plays a central role. For a matrix
$\mat{A}$, this means calculating a set of eigenvalues $\lambda_i$ and eigenvectors
$\vec{v}_i$, satisfying
\begin{equation}
  \mat{A} \vec{v}_i = \lambda_i \vec{v}_i.
\end{equation}
An example from classical mechanics or molecular science is the determination of
the principal axes of a solid object~\cite{Goldstein:ClassMech}. The author's
interest in the problem arises from the numerical computation of neutrino
oscillation probabilities in matter~\cite{Akhmedov:1999uz,Huber:2004ka,Huber:2007ji},
which requires the diagonalization of the Hamiltonian operator
\begin{equation}
  \mat{H} = \mat{U} \begin{pmatrix}
                      0  &                 &                 \\
                         & \Delta m_{21}^2 &                 \\
                         &                 & \Delta m_{31}^2
                       \end{pmatrix} \mat{U}^\dag
                     + \begin{pmatrix}
                         V &   &   \\
                           & 0 &   \\
                           &   & 0
                       \end{pmatrix}.
\end{equation}
Here, $\mat{U}$ is the leptonic mixing matrix, $\Delta m_{21}^2$ and $\Delta m_{31}^2$
are the differences of the squared neutrino masses, and $V$ is the MSW
(Mikheyev-Smirnov-Wolfenstein) Potential describing coherent forward
scattering in matter. If certain non-standard physics contributions
are considered, the MSW matrix can also contain more than one
non-zero entry~\cite{Roulet:1991sm}.

There exist many publicly available software packages for the calculation of matrix
eigensystems, e.g.\ LAPACK~\cite{Anderson:LAPACK}, the GNU Scientific
Library~\cite{Galassi:GSL}, or the Numerical Recipes algorithms~\cite{Press:NumRecip}.
These packages exhibit excellent accuracy, but being designed mainly for very large
matrices, they may produce a lot of computational overhead in the simple $3 \times 3$
case. This overhead comes partly from the algorithms themselves, and partly from the
implementational details.

In this letter, we will study the performance of several algorithms which
were optimized specifically for $3 \times 3$ matrices. We will discuss the
well-known Jacobi, QL and Cuppen algorithms, and compare their speed and
accuracy to that of a direct analytical calculation using Cardano's
formula for the eigenvalues, and vector cross products for the eigenvectors.
The application of Cardano's formula to the $3 \times 3$ eigenproblem
has been suggested previously in~\cite{Smith:1961}, and formulas for the
eigenvectors based on the computation of the Euler angles have been
presented in~\cite{Bojanczyk:1991}, 

The outline of the paper is as follows: In Secs.~\ref{sec:alg-iter}
and~\ref{sec:alg-noniter}, we will describe the mathematical background of the
considered algorithms as well as the most important implementational issues, and
discuss their numerical properties. In Sec.~\ref{sec:other}, we will briefly
mention some other algorithms capable of solving the $3 \times 3$ eigenproblem,
and give reasons why we do not consider them to be the optimal choice for such
small matrices. Our purely theoretical discussion will be complemented in
Sec.~\ref{sec:bench} by the presentation of detailed benchmark results. Finally,
we will draw our conclusions in Sec.~\ref{sec:conc}. The appendix contains two
alternative derivations of Cardano's formulas, and the documentation of our C and
Fortran code, which is available for download from
\texttt{http://www.mpi-hd.mpg.de/$\sim$globes/3x3/}.

\section{Iterative algorithms}
\label{sec:alg-iter}

\subsection{The Jacobi method}
\label{sec:jacobi}

One of the oldest methods for the diagonalization of an arbitrary symmetric or
hermitian $n \times n$ matrix $\mat{A}$ is the Jacobi algorithm. Discussions of this
algorithm can be found in~\cite{Press:NumRecip,Ralston:NumAnalysis,Fadeev:NumLinAlg,
Sueli:NumAnalysis,Forsythe:1960:CJM}. Its basic idea is to iteratively zero the
off-diagonal elements of $\mat{A}$ by unitary transformations of the form
\begin{equation}
  \mat{P}_{pq} =
    \begin{pmatrix}
      1 &        &                 &        &               &        &    \\
        & \ddots &                 &        &               &        &    \\
        &        &      c          & \cdots & s e^{i\alpha} &        &    \\
        &        &   \vdots        &   1    &   \vdots      &        &    \\
        &        & -s e^{-i\alpha} & \cdots &      c        &        &    \\
        &        &                 &        &               & \ddots &    \\
        &        &                 &        &               &        & 1
    \end{pmatrix}.
  \label{eq:pq-rot}
\end{equation}
The matrices $\mat{P}_{pq}$ differ from the unit matrix only in the $(pp)$,
$(pq)$, $(qp)$, and $(qq)$ elements; $c$ and $s$ are required to satisfy
$s^2 + c^2 = 1$, and can thus be expressed in the form
\begin{equation}
  \begin{aligned}
    c &= \cos\phi,  \\
    s &= \sin\phi
  \end{aligned}
\end{equation}
for some real angle $\phi$. The complex phase $\alpha$ is absent if $\mat{A}$
has only real entries. $\phi$ and $\alpha$ are chosen in such a way that the
similarity transformation
\begin{equation}
  \mat{A} \rightarrow \mat{P}_{pq}^\dag \, \mat{A} \, \mat{P}_{pq}
  \label{eq:jacobi-iter}
\end{equation}
eliminates the $(pq)$ element (and thus also the $(qp)$ element) of $\mat{A}$.
Of course it will in general become nonzero again in the next iteration, where
$p$ or $q$ is different, but one can show that the iteration~\eqref{eq:jacobi-iter}
converges to a diagonal matrix, with the eigenvalues of $\mat{A}$ along the diagonal,
if $p$ and $q$ cycle through the rows or columns of $\mat{A}$~\cite{Forsythe:1960:CJM}.
The normalized eigenvectors are given by the columns of the matrix
\begin{equation}
  \mat{Q} = \mat{P}_{p_1 q_1} \mat{P}_{p_2 q_2} \mat{P}_{p_3 q_3} \cdots.
\end{equation}

\subsection{The QR and QL algorithms}
\label{sec:ql}

The QR and QL algorithms are among the most widely used methods in large
scale linear algebra because they are very efficient and accurate although
their implementation is a bit more involved than that of the Jacobi
method~\cite{Press:NumRecip,Stoer:NumAnalysis,Fadeev:NumLinAlg}. They are
only competitive if applied to real symmetric tridiagonal matrices of the form
\begin{equation}
  \mat{T} =
    \begin{pmatrix}
      \times & \times &        &        &        &        \\
      \times & \times & \times &        &        &        \\
             & \times & \times & \times &        &        \\
             &        &        & \ddots &        &        \\
             &        &        & \times & \times & \times \\
             &        &        &        & \times & \times 
    \end{pmatrix}.
    \label{eq:tridiag}
\end{equation}
Therefore, as a preliminary step, $\mat{A}$ has to be brought to this form.

\subsubsection{Reduction of $\mat{A}$ to tridiagonal form}
\label{sec:reduction}

There are two main ways to accomplish the tridiagonalization:
The Givens method, which consists of successively applying plane rotations
of the form of Eq.~\eqref{eq:pq-rot} (in contrast to the Jacobi reduction to diagonal
form, the Givens reduction to tridiagonal form is non-iterative), and the
Householder method, which we will discuss here. A Householder transformation
is defined by the unitary transformation matrix
\begin{equation}
  \mat{P} = \mat{I} - \omega \vec{u} \vec{u}^\dagger
  \label{eq:Householder} 
\end{equation}
with
\begin{equation}
  \vec{u} = \vec{x} \mp |\vec{x}| \vec{e}_i
  \label{eq:Householder_u}
\end{equation}
and
\begin{equation}
  \omega =  \frac{1}{|\vec{u}|^2} \left( 1 + \frac{\vec{x}^\dag \vec{u}}
                                                  {\vec{u}^\dag \vec{x}} \right).
  \label{eq:Householder_omega}
\end{equation}
Here, $\vec{x}$ is arbitrary for the moment, and $\vec{e}_i$ denotes the $i$-th
unit vector. From a purely mathematical point of view, the choice of sign in
Eq.~\eqref{eq:Householder_u} is arbitrary, but in the actual implementation we
choose it to be equal to the sign of the real part of $x_i$ to avoid cancellation
errors.  $\mat{P}$ has the property that $\mat{P} \vec{x} \sim \vec{e}_i$ because
\begin{align}
  (\mat{I} - \omega \vec{u}\vec{u}^\dag) \vec{x} &=
      \vec{x} - \frac{\Big(1 + \frac{\vec{x}^\dag \vec{u}}{\vec{u}^\dag \vec{x}} \Big)
                    \vec{u}\vec{u}^\dag \vec{x}}
                       {2|\vec{x}|^2 \mp |\vec{x}|(x_i + x_i^*)}  \nonumber \\
   &\hspace{-1.7 cm}= \vec{x} - \frac{(\vec{u}^\dag\vec{x} + \vec{x}^\dag\vec{u})
                        (\vec{x} \mp |\vec{x}| \vec{e}_i)}
                       {2|\vec{x}|^2 \mp |\vec{x}|(x_i + x_i^*)}  \nonumber \\
   &\hspace{-1.7 cm}= \vec{x} - \frac{(|\vec{x}|^2 \mp |\vec{x}|x_i + 
                          |\vec{x}|^2 \mp |\vec{x}|x_i^*)
                        (\vec{x} \mp |\vec{x}| \vec{e}_i)}
                       {2|\vec{x}|^2 \mp |\vec{x}|(x_i + x_i^*)}  \nonumber \\
   &\hspace{-1.7 cm}= \pm |\vec{x}| \vec{e}_i.
\end{align}
This means, that if we choose $\vec{x}$ to contain the lower $n-1$ elements
of the first column of $\mat{A}$ and set $x_1 = 0$, $\vec{e}_i = \vec{e}_2$,
then
\begin{equation}
  \mat{P}_1 \, \mat{A} \, \mat{P}_1^\dag =
    \begin{pmatrix}
      \times & \times &        &        \\
      \times & \times & \cdots & \times \\
             & \vdots & \ddots & \vdots \\
             & \times & \cdots & \times
    \end{pmatrix}.
    \label{eq:HouseholderStep1}
\end{equation}
Note that the first row and the first column of this matrix are real even if
$\mat{A}$ is not. In the next step, $\vec{x}$ contains the lower $n-2$ elements
of the second column of $\mat{P}_1 \mat{A} \mat{P}_1^\dag$, $x_1 = x_2 = 0$,
and $\vec{e}_i = \vec{e}_3$, so that the second row (and column) is brought to
the desired form, while the first remains unchanged. This is repeated $n-1$
times, until $\mat{A}$ is fully tridiagonal and real.

For the actual implementation of the Householder method, we do not calculate
the matrix product $\mat{P} \mat{A} \mat{P}^\dag$ directly, but instead evaluate
\begin{align}
  \vec{p} &= \omega^* \mat{A} \vec{u},                 \label{eq:Householder_p} \\
  K       &= \frac{\omega}{2} \vec{u}^\dag \vec{p},    \label{eq:Householder_K} \\
  \vec{q} &= \vec{p} - K \vec{u}.                      \label{eq:Householder_q}
\end{align}
With these definitions, we have the final expression
\begin{align}
  \mat{P} \, \mat{A} \, \mat{P}^\dag &= \mat{P} (\mat{A} - \vec{p} \vec{u}^\dag) \nonumber\\
    &= \mat{A} - \vec{p} \vec{u}^\dag - \vec{u} \vec{p}^\dag
                + 2 K \vec{u} \vec{u}^\dag                                       \nonumber\\
    &= \mat{A} - \vec{q} \vec{u}^\dag - \vec{u} \vec{q}^\dag.
\end{align}
Note that in the last step we have made use of the fact that $K$ is real,
as can be seen from Eqs.~\eqref{eq:Householder_K} and~\eqref{eq:Householder_p},
and from the hermiticity of $\mat{A}$.

\subsubsection{The QL algorithm for real tridiagonal matrices}

The QL algorithm is based on the fact that any real matrix can be decomposed
into an orthogonal matrix $\mat{Q}$ and a lower triangular matrix $\mat{L}$
according to
\begin{equation}
  \mat{A} = \mat{Q} \mat{L}.
\end{equation}
Equivalently, one could also start from a decomposition of the form
$\mat{A} = \mat{Q} \mat{R}$, with $\mat{R}$ being upper triangular,
to obtain the QR algorithm, which has similar properties.
For tridiagonal $\mat{A}$, the QL decomposition is most efficiently calculated
by a sequence of plane rotations of the form of Eq.~\eqref{eq:pq-rot}.
The iteration prescription is
\begin{equation}
  \mat{A} \rightarrow \mat{Q}^T \, \mat{A} \, \mat{Q},
  \label{eq:qr-iter}
\end{equation}
but to accelerate convergence it is advantageous to use the method of \emph{shifting},
which means decomposing \mbox{$\mat{A} - k \mat{I}$} instead of $\mat{A}$. In each
step, $k$ is chosen in such a way that the convergence to zero of the uppermost
non-zero off-diagonal element of $\mat{A}$ is maximized (see~\cite{Stoer:NumAnalysis}
for a discussion of the shifting strategy and for corresponding convergence theorems).
Since in practice, the subtraction of $k$ from the diagonal elements of $\mat{A}$
can introduce large numerical errors, the most widely used form of the QL algorithm
is one with \emph{implicit shifting}, where not all of these differences need to be
evaluated explicitly, although the method is mathematically equivalent to the ordinary
QL algorithm with shifting.

\subsection{Efficiency and accuracy of the Jacobi and QL algorithms}
\label{sec:JacobiEff}

As we have mentioned before, one of the main benefits of the QL algorithm is
its efficiency: For matrices of large dimension $n$, it requires $\approx 30 n^2$
floating point operations (combined multiply/add) if only the eigenvalues
are to be computed, or $\approx 6 n^3$ operations if also the complex eigenvectors
are desired~\cite{Press:NumRecip}. Additionally, the preceding complex Householder
transformation requires $8 n^3/3$ resp.\ $16 n^3/3$ operations. In contrast, the
complex Jacobi algorithm takes about $3 n^2$ to $5 n^2$ complex Jacobi rotations,
each of which involves $12 n$ operations for the eigenvalues, or $24 n$ operations
for the complete eigensystem. Therefore, the total workload is
$\approx 35 n^3$ -- $60 n^3$ resp.\ $70 n^3$ -- $120 n^3$ operations.

For the small matrices considered here, these estimates are not reliable.
In particular, the QL method suffers from the fact that the first few eigenvalues
require more iterations than the later ones, since for these the corresponding
off-diagonal elements have already been brought close to zero in the preceding
steps. Furthermore, the operations taking place before and after the innermost
loops will play a role for small matrices. Since these are more complicated
for QL than for Jacobi, they will give an additional penalty to QL.
For these reasons we expect the performance bonus of QL over Jacobi
to be smaller for $3 \times 3$ matrices than for larger problems.

The numerical properties of both iterative methods are independent of the
matrix size and have been studied in great detail by others, so we will only give
a brief overview here. For real symmetric positive definite matrices, Demmel
and Veseli\'{c} have shown that the Jacobi method is more accurate than the QL
algorithm~\cite{Demmel:1989,Demmel:1992}. In particular, if the eigenvalues
are distributed over many orders of magnitude, QL may become inaccurate for
the small eigenvalues. In the example given in~\cite{Demmel:1992}, the extremely
fast convergence of this algorithm is also its main weakness: In the attempt
to bring the lowest diagonal element close to the smallest eigenvalue of the matrix,
a difference of two almost equal numbers has to be taken.

If the requirement of positive definiteness is omitted, one can also
find matrices for which QL is more accurate than Jacobi~\cite{Stewart:1995}.

\section{Non-iterative algorithms}
\label{sec:alg-noniter}

\subsection{Direct analytical calculation of the eigenvalues}
\label{sec:alg-noniter-values}

For $3 \times 3$ matrices, the fastest way to calculate the eigenvalues is by directly
solving the characteristic equation
\begin{equation}
  P(\lambda) = |\mat{A} - \lambda \mat{I}| = 0
  \label{eq:chareq}
\end{equation}
If we write
\begin{equation}
  \mat{A} = \begin{pmatrix}
              a_{11}   & a_{12}   & a_{13}  \\
              a_{12}^* & a_{22}   & a_{23}  \\
              a_{13}^* & a_{23}^* & a_{33}
            \end{pmatrix},
\end{equation}
Eq.~\eqref{eq:chareq} takes the form
\begin{equation}
  P(\lambda) = \lambda^3 + c_2 \lambda^2 + c_1 \lambda + c_0 = 0
  \label{eq:charpoly}
\end{equation}
with the coefficients
\begin{align}
  c_2 &= -a_{11} - a_{22} - a_{33},                                
    \label{eq:c2} \\
  c_1 &= a_{11} a_{22} + a_{11} a_{33} + a_{22} a_{33}             \nonumber\\
      &  \hspace{2 cm}  - |a_{12}|^2 - |a_{13}|^2 - |a_{23}|^2,
    \label{eq:c1} \\
  c_0 &= a_{11} |a_{23}|^2 + a_{22} |a_{13}|^2 + a_{33} |a_{12}|^2 \nonumber\\
      &  \hspace{1 cm}  - a_{11} a_{22} a_{33}
                        - 2 \, \Re (a_{13}^* a_{12} a_{23}).
    \label{eq:c0}
\end{align}
To solve Eq.~\eqref{eq:charpoly}, we follow the method proposed by del~Ferro,
Tartaglia, and Cardano in the 16$^\textrm{th}$ century~\cite{Cardano:1545}. In
Appendix~\ref{sec:cardano-alt} we will discuss two alternative approaches
and show that they lead to the same algorithm as Cardano's method if numerical
considerations are taken into account.

Cardano's method requires first transforming Eq.~\eqref{eq:charpoly} to the form
\begin{equation}
  x^3 - 3x = t,
  \label{eq:charpoly-reduced}
\end{equation}
by defining
\begin{align}
  p &= c_2^2 - 3 c_1,                                     \\
  q &= -\tfrac{27}{2} c_0 - c_2^3 + \tfrac{9}{2} c_2 c_1, \\
  t &= 2 p^{-3/2} q,                                      \\
  x &= \tfrac{3}{\sqrt{p}} (\lambda + \tfrac{1}{3} c_2).
  \label{eq:tpq}
\end{align}
It is easy to check that a solution to Eq.~\eqref{eq:charpoly-reduced} is
then given by
\begin{equation}
  x = \frac{1}{u} + u,
\end{equation}
with
\begin{equation}
  u = \sqrt[3]{\frac{t}{2} \pm \sqrt{\frac{t^2}{4} - 1}}.
\end{equation}
There is a sixfold ambiguity in $u$ (two choices for the sign of the square root
and three solutions for the complex cube root), but it reduces to the
expected threefold ambiguity in $x$.

To achieve optimal performance, we would like to avoid complex arithmetic
as far as possible. Therefore, we will now show that $\sqrt{p}$, and thus
$t$, are always real. We know from linear algebra that the characteristic
polynomial $P(\lambda)$ of the hermitian matrix $\mat{A}$ must have three
real roots, which is only possible if the stationary points of $P(\lambda)$,
$\tilde{\lambda}_{1/2} = -\tfrac{1}{3} c_2 \pm \tfrac{1}{3} \sqrt{c_2^2 - 3c_1} =
-\tfrac{1}{3} c_2 \pm \tfrac{1}{3} \sqrt{p}$ are real. Since $c_2$ is real, this
implies that also $\sqrt{p}$ must be real.

Furthermore, from the same argument, we have the requirement that
$P(\tilde{\lambda}_1) \geq 0 \geq P(\tilde{\lambda}_2)$, which in turn
implies that $-2 \leq t \leq 2$. Therefore, $\sqrt{t^2/4 - 1}$ is always
purely imaginary, and from this it is easy to see that $|u| = 1$.
Therefore we can write $u = e^{i \phi}$, with
\begin{align}
  \phi &= \tfrac{1}{3} \arctan \frac{\sqrt{|t^2/4 - 1|}}{t/2}
        = \tfrac{1}{3} \arctan \frac{\sqrt{p^3 - q^2}}{q} \nonumber \\
       &= \tfrac{1}{3} \arctan \frac{\sqrt{27 \big[ \tfrac{1}{4} c_1^2 (p - c_1)
                            + c_0 (q + \tfrac{27}{4} c_0) \big]}}{q}.
  \label{eq:phi}
\end{align}
The last step is necessary to improve the numerical accuracy, which
would suffer dramatically if we computed the difference $p^3 - q^2$
directly.

When evaluating Eq.~\eqref{eq:phi}, care must be taken to correctly
resolve the ambiguity in the $\arctan$ by taking into account the
sign of $q$: For $q > 0$, the solution must lie in the first
quadrant, for $q < 0$ it must be located in the second. In contrast
to this, solutions differing by multiples of $2\pi$ are equivalent,
so $x$ can take three different values,
\begin{equation}
  \begin{aligned}
  x_1 &= 2 \cos\phi,                               \\
  x_2 &= 2 \cos \Big(\phi + \frac{2\pi}{3} \Big)
       = -\cos\phi - \sqrt{3} \sin\phi,            \\
  x_3 &= 2 \cos \Big(\phi - \frac{2\pi}{3} \Big)
       = -\cos\phi + \sqrt{3} \sin\phi.
  \end{aligned}
  \label{eq:cardano}
\end{equation}
These correspond to the three eigenvalues of $\mat{A}$:
\begin{equation}
  \lambda_i = \tfrac{\sqrt{p}}{3} x_i - \tfrac{1}{3} c_2.
  \label{eq:lambda_i}
\end{equation}
Similar formulas have been derived previously in~\cite{Smith:1961}.

The most expensive steps of Cardano's algorithm are the evaluations of the
trigonometric functions. Nevertheless, the method is extremely fast, and will
therefore be the best choice for many practical problems. However, from
Eq.~\eqref{eq:lambda_i} we can see that it becomes unstable for matrices with
largely different eigenvalues: In general, $c_2$ is of the order of the
largest eigenvalue $\lambda_{\max}$. Therefore, in order to obtain the
smaller eigenvalues, considerable cancellation between $\tfrac{\sqrt{p}}{3} x_i$
and $\tfrac{1}{3} c_2$ must occur, which can yield large errors and
is very susceptible to tiny relative errors in the calculation of $c_2$.

In general, the roots of Eq.~\eqref{eq:charpoly} can be very sensitive to
small errors in the coefficients which might arise due to cancellations
in Eqs.~\eqref{eq:c2} -- \eqref{eq:c0}.

If $\varepsilon$ is the machine precision, we can estimate the absolute
accuracy of the eigenvalues to be of
$\mathcal{O}(\varepsilon \lambda_{\max})$,
which may imply a significant loss of \emph{relative} accuracy for
the small eigenvalues.

Consider for example the matrix
\begin{equation}
  \begin{pmatrix}
    10^{40} & 10^{19} & 10^{19} \\
    10^{19} & 10^{20} & 10^{9}  \\
    10^{19} & 10^{9}  & 1
  \end{pmatrix},
  \label{eq:testmatrix1}
\end{equation}
which has the (approximate) eigenvalues $10^{40}$, $10^{20}$, and $1$.
However, Cardano's method yields $10^{40}$, $5 \cdot 10^{19}$, and
$-5 \cdot 10^{19}$. Note that for the matrix~\eqref{eq:testmatrix1},
also the QL algorithm has problems and delivers one negative eigenvalue.
Only Jacobi converges to a reasonable relative accuracy.
See~\cite{Demmel:1989} for a discussion of this.

\subsection{Direct analytical calculation of the eigenvectors}
\label{sec:alg-noniter-vectors}

Once the eigenvalues $\lambda_i$ of $\mat{A}$ have been computed, the
eigenvectors $\vec{v}_i$ can be calculated very efficiently by using
vector cross products, which are a unique tool in three-dimensional space.

The $\vec{v}_i$ satisfy by definition
\begin{equation}
  (\mat{A} - \lambda_i \mat{I}) \cdot \vec{v}_i = 0.
\end{equation}
Taking the hermitian conjugate of this equation and multiplying it with an
arbitrary vector $\vec{x} \in \mathbb{C}^3$, we obtain
\begin{equation}
  \vec{v}_i^\dagger \cdot (\mat{A} - \lambda_i \mat{I}) \cdot \vec{x} = 0.
\end{equation}
In particular, if $\vec{x}$ is the $j$-th unit vector $\vec{e}_j$,
this becomes
\begin{equation}
  \vec{v}_i^\dagger \cdot (\vec{A}^j - \lambda_i \vec{e}_j) = 0
      \hspace{1 cm} \forall \ j,
\end{equation}
where $\vec{A}^j$ denotes the $j$-th column of $\mat{A}$.
Consequently, as long as $\vec{A}^1 - \lambda_i \vec{e}_1$ and
$\vec{A}^2 - \lambda_i \vec{e}_2$ are linearly independent, we have
\begin{equation}
  \vec{v}_i = [ (\vec{A}^1 - \lambda_i \vec{e}_1) \times
                   (\vec{A}^2 - \lambda_i \vec{e}_2) ]^*.
  \label{eq:crossprod}
\end{equation}
In the special case that
$\vec{A}^1 - \lambda_i \vec{e}_1 = \mu (\vec{A}^2 - \lambda_i \vec{e}_2)$,
$\vec{v}_i$ is immediately given by
\begin{equation}
  \vec{v}_i = \frac{1}{\sqrt{1 + |\mu|^2}} \begin{pmatrix}
                                             1  \\  -\mu  \\  0
                                           \end{pmatrix}.
\end{equation}

When implementing the above procedure, care must be taken if there is a degenerate
eigenvalue because in this case, the algorithm will only find one of the two
corresponding eigenvectors. Therefore, if we detect a degenerate eigenvalue,
say $\lambda_1 = \lambda_2$, we calculate the second eigenvector as the cross
product of $\vec{v}_1$ with one of the columns of $\mat{A} - \lambda_1 \mat{I}$.
In principle, this alternative formula would also work for non-degenerate
eigenvalues, but we try to avoid it as far as possible because it abets the
propagation of errors. On the other hand, we have to keep in mind that the
test for degeneracy may fail if the eigenvalues have been calculated too
inaccurately. If this happens, the algorithm will deliver wrong results.

The calculation of the third eigenvalue can be greatly accelerated by using
the formula $\vec{v}_3^\dag = \vec{v}_1 \times \vec{v}_2$. Of course, this is
again vulnerable to error propagation, but it turns out that this is usually
tolerable.

For many practical purposes that require only moderate accuracy, the vector
product algorithm is the method of choice because it is considerably faster
than all other approaches. However, the limitations to its accuracy need to
be kept in mind. First, the eigenvectors suffer from errors in the eigenvalues.
Under certain circumstances, these errors can even be greatly enhanced by the
algorithm. For example, the matrix
\begin{equation}
  \begin{pmatrix}
    10^{20} & 10^{9}  & 10^{9} \\
    10^{9}  & 10^{20} & 10^{9} \\
    10^{9}  & 10^{9}  & 1
  \end{pmatrix},
  \label{eq:testmatrix2}
\end{equation}
has the approximate eigenvalues $(1 + 10^{-11}) \cdot 10^{20}$,
$(1 - 10^{-11}) \cdot 10^{20}$, and $0.98$. The corresponding eigenvectors
are approximately
\begin{equation}
  \vec{v}_1 = \begin{pmatrix}
                1/\sqrt{2} \\
                1/\sqrt{2} \\
                0
              \end{pmatrix}, \hspace{0.4 cm}
  \vec{v}_2 = \begin{pmatrix}
                1/\sqrt{2} \\
                -1/\sqrt{2} \\
                0
              \end{pmatrix}, \hspace{0.4 cm}
  \vec{v}_3 = \begin{pmatrix}
                10^{-11} \\
                10^{-11} \\
                1
              \end{pmatrix}.
\end{equation}
If we erroneously start the vector product algorithm with the approximate
eigenvalues $10^{20}$, $10^{20}$, and $0.98$, the error that is introduced
when subtracting $\lambda_1$ from the diagonal elements is of order
$\mathcal{O}(10^9)$ and thus comparable to the off-diagonal elements.
Consequently, the calculated eigenvectors
\begin{equation}
  \vec{v}_1 = \begin{pmatrix}
                 1/\sqrt{3} \\
                 1/\sqrt{3} \\
                -1/\sqrt{3}
              \end{pmatrix}, \hspace{0.4 cm}
  \vec{v}_2 = \begin{pmatrix}
                 2 / \sqrt{6} \\
                -1 / \sqrt{6} \\
                 1 / \sqrt{6}
              \end{pmatrix}, \hspace{0.4 cm}
  \vec{v}_3 = \begin{pmatrix}
                0        \\
                1/\sqrt{2} \\
                1/\sqrt{2} \\
              \end{pmatrix}
\end{equation}
are completely wrong.

Another flaw of the vector product algorithm is the fact that the subtractions
$(\vec{A}^j - \lambda_i \vec{e}_j)$ and the subtractions in the
evaluation of the cross products are very prone to cancellation errors.

\subsection{A hybrid algorithm}
\label{sec:alg-noniter-hybrid}

To circumvent the cases where the cross product method fails or becomes too
inaccurate, we have devised a hybrid algorithm, which uses the analytical
calculations from Secs.~\ref{sec:alg-noniter-values} and~\ref{sec:alg-noniter-vectors}
as the default branch, but falls back to QL if this procedure is estimated to
be too error-prone. The condition for the fallback is
\begin{equation}
  || \vec{v}_i ||^2 \leq 2^8 \, \epsilon \Lambda^2
  \label{eq:hybrid-cond}
\end{equation}
Here, $\vec{v}_i$ is the analytically calculated and yet unnormalized eigenvector
from Eq.~\eqref{eq:crossprod}, $\epsilon$ is the machine precision, 
$\Lambda = {\max}(\lambda_{\max}^2, \lambda_{\max}^{\phantom 2})$ is an estimate
for the largest number appearing in the problem, and $2^8$ is introduced as a safety
factor.

Since in typical problems only a small fraction of matrices will fulfill
condition~\eqref{eq:hybrid-cond}, the hybrid approach can be expected to retain the
efficiency of the analytical algorithm. In reality, it is even slightly faster
because fewer conditional branches are used.

\subsection{Cuppen's Divide and Conquer algorithm}
\label{sec:cuppen}

In recent years, the ``Divide and Conquer'' paradigm for symmetric
eigenproblems has received considerable attention. The idea was originally
invented by Cuppen~\cite{Cuppen:1981}, and current implementations are faster
than the QL method for large matrices~\cite{Anderson:LAPACK}. One can estimate,
that for $3 \times 3$ matrices, a divide and conquer strategy might also be beneficial,
because it means splitting the problem into a trivial $1 \times 1$ and an
analytically accessible $2 \times 2$ problem.

However, let us first discuss Cuppen's algorithm in its general form for $n \times n$
matrices. As a preliminary step, the matrix has to be reduced to symmetric
tridiagonal form, as discussed in Sec.~\ref{sec:reduction}. The resulting
matrix $\mat{T}$ is then split up in the form
\begin{equation}
  \mat{T} = \begin{pmatrix}
              \mat{T}_1 & 0   \\
              0         & \mat{T}_2
            \end{pmatrix}  +  \mat{H},
\end{equation}
where $\mat{T}_1$ and $\mat{T}_2$ are again tridiagonal, and $\mat{H}$ is a very
simple rank~1 matrix, e.g.
\begin{equation}
  \mat{H} = \begin{pmatrix}
              \mat{0} &       &       &         \\
                      & \beta & \beta &         \\
                      & \beta & \beta &         \\
                      &       &       & \mat{0} \\
            \end{pmatrix}.
\end{equation}
Then, the smaller matrices $\mat{T}_1$ and $\mat{T}_2$ are brought to the diagonal form
$\mat{T}_i = \mat{Q}_i \mat{D}_i \mat{Q}_i^T$, so that $\mat{T}$ becomes
\begin{align}
  \mat{T} &= \begin{pmatrix}
               \mat{Q}_1 \mat{D}_1 \mat{Q}_1^T & 0                                 \\
               0                               & \mat{Q}_2 \mat{D}_2 \mat{Q}_2^T
             \end{pmatrix} + \mat{H}  \nonumber\\ &=
             \begin{pmatrix}
               \mat{Q}_1 & 0   \\
               0         & \mat{Q}_2
             \end{pmatrix} \left[
             \begin{pmatrix}
               \mat{D}_1 & 0   \\
               0         & \mat{D}_2
             \end{pmatrix} + \mat{H}^\prime \right]
             \begin{pmatrix}
               \mat{Q}_1^T & 0   \\
               0           & \mat{Q}_2^T
             \end{pmatrix}.
\end{align}
Here, $\mat{H}^\prime = \vec{z} \, \vec{z}^T$ is another rank~1 matrix with a
generating vector $\vec{z}$ consisting of the last row of $\mat{Q}_1$ and the
first row of $\mat{Q}_2$, both multiplied with $\beta$. The remaining problem
is to find an eigenvalue $\lambda$ and an eigenvector $\vec{v}$, satisfying
\begin{equation}
  \left[ \begin{pmatrix}
           \mat{D}_1 & 0   \\
           0         & \mat{D}_2
         \end{pmatrix}
         + \vec{z} \, \vec{z}^T - \lambda \mat{I} \right] \vec{v} = 0.
\end{equation}
By multiplying this equation from the left with
\mbox{$\vec{z}^T \cdot \diag((\mat{D}_1 - \lambda \mat{I})^{-1}, (\mat{D}_2 - \lambda \mat{I})^{-1})$}
and dividing off the scalar $\vec{z}^T \vec{v}$, we obtain the characteristic
equation in the form
\begin{equation}
  1 + \sum_i \frac{z_i^2}{d_i - \lambda} = 0,
  \label{eq:Cuppen-secular}
\end{equation}
where the $d_i$ are the diagonal entries of $\mat{D}_1$ and $\mat{D}_2$.

There are several obvious strategies for solving this equation in the $3 \times 3$
case: First, by multiplying out the denominators we obtain a third degree polynomial
which can be solved exactly as discussed in Sec.~\ref{sec:alg-noniter-values}. This is
very fast, but we have seen that it can be numerically unstable. Second, one could
apply the classical Newton-Raphson iteration, which is fast and accurate in the
vicinity of the root, but may fail altogether if a bad starting value is chosen.
Finally, the fact that the roots of Eq.~\eqref{eq:Cuppen-secular} are known to be
separated by the singularities $d_1$, $d_2$ and $d_3$ suggests the usage of a
bisection method to obtain the eigenvalues very accurately, but with slow convergence.

To get the optimum results, we apply Cardano's analytical method to get estimates
for the roots, and refine them with a hybrid Newton-Raphson/Bisection
method based on~\cite{Press:NumRecip}. This method usually takes Newton-Raphson
steps, but if the convergence gets too slow or if Newton-Raphson runs out of
the bracketing interval, it falls back to bisection.

The elements $\tilde{v}_{ij}$ of the eigenvectors $\vec{\tilde{v}}_i$ of
$\mathrm{diag}(\mat{D}_1, \mat{D}_2) + \mat{H}^\prime$ are then obtained by the
simple formula
\begin{equation}
  \tilde{v}_{ij} = \frac{z_j}{d_j - \lambda_i}
  \label{eq:Cuppen-vectors}
\end{equation}
and just need to be transformed back to the original basis by undoing the
transformations $\mat{Q}_1$, $\mat{Q}_2$, and the tridiagonalization.

If implemented carefully, Cuppen's method can reach an accuracy comparable
to that of the QL method. A major issue is the calculation of the differences
$d_i - \lambda_j$ in the evaluation of the characteristic equation,
Eq.~\eqref{eq:Cuppen-secular}, and in the calculation of the eigenvectors,
Eq.~\eqref{eq:Cuppen-vectors}. To keep the errors as small as possible
when solving for the eigenvalue $\lambda_j$, we subtract our initial estimate
for $\lambda_j$ from all $d_i$ before starting the iteration. This ensures that
the thus transformed eigenvalue is very close to zero and therefore small
compared to the $d_i$.

As we have mentioned before, the Divide and Conquer algorithm is faster than
the QL method for large matrices. It also required $\mathcal{O}(n^3)$ operations,
but since the expensive steps --- the reduction to tridiagonal form and the
back-transformation of the eigenvectors --- are both outside the iterative loops,
the coefficient of $n^3$ is significantly smaller. For the small matrices that
we are considering, however, the most expensive part is solving the characteristic
equation. Furthermore, many conditional branches are required to implement
necessary case differentiations, to avoid divisions by zero, and to handle
special cases like multiple eigenvalues. Therefore, we expect the algorithm
to be about as fast as the QL method.

It is of course possible to reduce the calculational effort at the expense
of reducing the accuracy and stability of the algorithm, but it will always be
slower than Cardano's method combined with the vector product algorithm.

\section{Other algorithms}
\label{sec:other}

Apart from the Jacobi, QL, Cuppen, and vector product algorithms there are
several other methods for finding the eigensystems of symmetric matrices. We
will briefly outline some of them here, and give reasons why they are
inappropriate for $3 \times 3$ problems.

\subsection{Iterative root finding methods}

In order to avoid the instabilities of Cardano's method which were discussed
in Sec.~\ref{sec:alg-noniter-values}, one can use an iterative root finding
method to solve the characteristic equation. Root bracketing algorithms like
classical bisection or the Dekker-Brent method start with an interval
which is known to contain the root. This interval is then iteratively narrowed
until the root is known with the desired accuracy. Their speed of convergence
is fair, but they are usually superseded by the Newton-Raphson method which
follows the gradient of the function until it finds the root. However, the
Newton-Raphson algorithm is not guaranteed to converge in all cases.

Although these problems can partly be circumvented by using a hybrid method
like the one discussed in Sec.~\ref{sec:cuppen}, iterative root finders are
still unable to find multiple roots, and these special cases would have to be
treated separately. Furthermore, the accuracy is limited by the accuracy with
which the characteristic polynomial can be evaluated. As we have already
mentioned in Sec.~\ref{sec:alg-noniter-values}, this can be spoiled by
cancellations in the calculation of the coefficients $c_0$, $c_1$, and $c_2$.

\subsection{Inverse iteration}

Inverse iteration is a powerful tool for finding eigenvectors and improving
the accuracy of eigenvalues. The method starts with some approximation
$\tilde{\lambda}_i$ for the desired eigenvalue $\lambda_i$, and a random vector
$\vec{b}$. One then solves the linear equation
\begin{equation}
  (\mat{A} - \tilde{\lambda}_i \mat{I}) \cdot \vec{\tilde{v}}_i = \vec{b}
\end{equation}
to find and approximation $\vec{\tilde{v}}_i$ for the eigenvector $\vec{v}_i$.
An improved estimate for $\lambda_i$ is calculated by using the formula
\begin{equation}
  (\mat{A} - \tilde{\lambda}_i \mat{I}) \cdot \vec{\tilde{v}}_i
      \approx (\lambda_i - \tilde{\lambda}_i) \cdot \vec{\tilde{v}}_i.
\end{equation}
We estimate that inverse iteration is impractical for small matrices because
there are many special cases that need to be detected and handled separately.
This would slow the computation down significantly.

\subsection{Vectorization}

In a situation where a large number $N$ of small matrices needs to be diagonalized,
and all these matrices are available at the same time, it may be advantageous
to vectorize the algorithm, i.e.\ to make the loop from $1$ to $N$ the
innermost loop~\footnote{We thank Charles van Loan for drawing our attention to
this possibility.}. This makes consecutive operations independent of each other
(because they affect different matrices), and allows them to be pipelined and
executed very efficiently.

A detailed discussion of this approach is beyond the scope of the present work,
but our estimate is that, as long as only the eigenvalues are to be computed, a
vectorization of Cardano's method would be most beneficial, because this
method requires only few performance-limiting conditional branches, so that
the number of processor pipeline stalls is reduced to a minimum. However,
the accuracy limitations discussed above would still apply in the vectorized version.

If we want to calculate the full eigensystem, a vectorized vector product method
can only give a limited performance bonus because in the calculation of the
vector products, many special cases can arise which need to be detected and
treated separately. This renders efficient vectorization impossible. The same
is true for Cuppen's Divide and Conquer algorithm. On the other hand, the
iterative methods are problematic if the required number of iterations is not
approximately the same for all matrices. Then, only the first few iterations
can be vectorized efficiently. Afterwards, matrices for which the algorithm has
not converged yet need to be treated separately.

\section{Benchmark results}
\label{sec:bench}

In this section we report on the computational performance and on the numerical
accuracy of the above algorithms. For the iterative methods we use implementations
which are similar to those discussed in~\cite{Press:NumRecip}. Additionally, we
study the LAPACK implementation of the QL/QR algorithm~\cite{Anderson:LAPACK} and
the QL routine from the GNU Scientific Library~\cite{Galassi:GSL}. For the
analytical methods we use our own implementations. Some ideas in our Cuppen
routine are based on ideas realized in LAPACK.

Note that we do not show results for the LAPACK implementation of a Divide and
Conquer algorithm (routine \texttt{ZHEEVD}) because this algorithm falls back to QL
for small matrices ($n < 25$) and would therefore not yield anything new. We also
neglect the new LAPACK routine \texttt{ZHEEVR} because for the $3 \times 3$ problems
considered here it is significantly slower than the other algorithms.

We have implemented our code in C and Fortran, but here we will only discuss
results for the Fortran version. We have checked that the C code yields a similar
numerical accuracy, but is about 10\% slower. This performance deficit can be
ascribed to the fact that C code is harder to optimize for the compiler.

Our code has been compiled with the GNU compiler, using the standard optimization
flag \texttt{-O3}. We did not use any further compiler optimizations although
we are aware of the fact that it is possible to increase the execution speed
by about 10\% if options like \texttt{-ffast-math} are used to allow optimizations
that violate the IEEE~754 standard for floating point arithmetic.

Our numerical tests were conducted in double precision arithmetic (64 bit,
15 -- 16 decimal digits) on an AMD Opteron~250 (64-bit, 2.4~GHz)
system running Linux. To maximize the LAPACK performance on this system, we used
the highly optimized AMD Core Math Library for the corresponding tests. Note that
on some platforms, in particular on Intel and AMD desktop processors, you may
obtain a higher numerical accuracy than is reported here because these processors
internally use 80 bit wide floating point registers.

To measure the accuracy of the calculated eigensystems we use three different
benchmarks:
\begin{itemize}
  \item The relative difference of the eigenvalue $\tilde{\lambda}$ and the
    corresponding result of the LAPACK QL routine \texttt{ZHEEV},
    $\tilde{\lambda}^\mathrm{LAPACK}$:
    \begin{equation}
      \Delta_1 \equiv \left|
                   \frac{\tilde{\lambda} - \tilde{\lambda}^\mathrm{LAPACK}}
                        {\tilde{\lambda}^\mathrm{LAPACK}} \right|.
      \label{eq:Delta1}
    \end{equation}
    We use the LAPACK QL implementation as a reference because it is very well tested
    and stable.
  \item The relative difference of the eigenvector $\tilde{\vec{v}}$
    and the corresponding LAPACK result, $\tilde{\vec{v}}^\mathrm{LAPACK}$:
    \begin{equation}
      \Delta_2 \equiv 
        \frac{\| \tilde{\vec{v}} - \tilde{\vec{v}}^\mathrm{LAPACK} \|_2}
             {\| \tilde{\vec{v}}^\mathrm{LAPACK} \|_2},
      \label{eq:Delta2}
    \end{equation}
    where $\| \cdot \|_2$ denotes the Euclidean norm. This definition of $\Delta_2$
    is of course not meaningful for matrices with degenerate eigenvalues because
    for these, different algorithms might find different bases for the
    multidimensional eigenspaces. Even for non-degenerate eigenvalues, $\Delta_2$ is
    only meaningful if the same phase convention is chosen for $\tilde{\vec{v}}$ and
    $\tilde{\vec{v}}^\mathrm{LAPACK}$. Therefore, when computing $\Delta_2$, we
    reject matrices with degenerate eigenvalues, and for all others we re-phase the
    eigenvectors in such a way that the component which has the largest modulus in
    $\tilde{\vec{v}}^\mathrm{LAPACK}$ is real.
  \item The deviation of the eigenvalue/eigenvector pair
    $(\tilde{\lambda}, \tilde{\vec{v}})$ from its defining property
    $\mat{A} \vec{v} = \lambda \vec{v}$:
    \begin{equation}
      \Delta_3 \equiv 
        \frac{\| \mat{A} \tilde{\vec{v}} - \tilde{\lambda} \tilde{\vec{v}} \|_2}
             {\| \tilde{\lambda} \tilde{\vec{v}} \|}.
      \label{eq:Delta3}
    \end{equation}
\end{itemize}
Note that for the algorithms considered here, $\| \tilde{\vec{v}} \|_2 = 1$,
so the definitions of $\Delta_2$ and $\Delta_3$ can be slightly simplified.

We have first verified the correctness of the algorithms by diagonalizing $10^7$
random hermitian resp.\ symmetric matrices with integer entries from the
interval $[-10, 10]$ and (automatedly) checking that $\Delta_1$, $\Delta_2$ and
$\Delta_3$ were as small as expected. We have repeated this test with
logarithmically distributed integer entries from the interval $[0, 10^{10}]$.
Such matrices tend to have largely different eigenvalues and are therefore a
special challenge to the algorithms.

For the results that we are going to present here, we have used similar tests, but
we have allowed the matrix entries to be real, which corresponds more closely
to what is found in actual scientific problems. Furthermore, for the
logarithmically distributed case we have changed the interval to $[10^{-5}, 10^{5}]$.

\subsection{Performance}

The results of our timing tests are summarized in Table~\ref{tab:speed}.
The most important observation from this table is the fact that the standard
libraries are slower by a substantial factor than the specialized algorithms. This
is due to the fact that in both LAPACK and the GSL, only the innermost loops, which
determine the performance for large scale problems, are optimized. Outside these
loops, the libraries spend a lot of time evaluating their parameters and preparing
data structures. LAPACK QL additionally takes some time to decide at runtime whether
a QR or QL algorithm is more favorable.

\begin{table}
  \centering
  \textbf{Real symmetric matrices}

  \vspace{0.2 cm}
  \begin{ruledtabular}
  \begin{tabular}{l|c|c|c|c}
    Algorithm         & \multicolumn{2}{c|}{Eigenvalues}
                                 &  \multicolumn{2}{c}{Eigenvectors} \\ \hline
                      & Lin.  & Log.  & Lin.  & Log.  \\ \hline

     Jacobi           &  12.0 &  10.0 &  12.9 &  10.9 \\
     QL               &   8.3 &   6.3 &   9.0 &   6.9 \\
     Cuppen           &   6.6 &   8.7 &   9.3 &  11.5 \\
     GSL              &  14.1 &  11.6 &  18.8 &  15.7 \\
     LAPACK QL        &  21.5 &  19.3 &  49.9 &  45.2 \\
     Analytical       &   2.9 &   3.0 &   4.0 &   4.1 \\
     Hybrid           &   3.0 &   3.8 &   3.4 &   4.5 \\
  \end{tabular}
  \end{ruledtabular}
  
  \vspace{0.5 cm}
  \textbf{Complex hermitian matrices}

  \vspace{0.2 cm}
  \begin{ruledtabular}
  \begin{tabular}{l|c|c|c|c}
    Algorithm         & \multicolumn{2}{c|}{Eigenvalues}
                                 &  \multicolumn{2}{c}{Eigenvectors} \\ \hline
                      & Lin.  & Log.  & Lin.  & Log.  \\ \hline

     Jacobi           &  17.0 &  15.4 &  21.0 &  18.4 \\
     QL               &  10.8 &   8.9 &  13.5 &  11.2 \\
     Cuppen           &   8.3 &   9.8 &  12.5 &  14.1 \\
     GSL              &  20.4 &  17.8 &  38.2 &  32.5 \\
     LAPACK QL        &  29.6 &  26.7 &  61.9 &  57.6 \\
     Analytical       &   3.5 &   3.7 &   5.9 &   6.0 \\
     Hybrid           &   3.7 &   3.6 &   4.8 &   5.1 \\
  \end{tabular}
  \end{ruledtabular}

  \caption{Performance of different algorithms for calculating the eigenvalues
    and eigenvectors of symmetric or hermitian $3 \times 3$ matrices. We show the
    running times in seconds for the calculation of only the eigenvalues (left)
    and of the complete eigensystems (right) of $10^7$ random matrices. Furthermore,
    we compare the cases of linearly and logarithmically distributed matrix entries.
    For the scenarios with real matrices, specialized versions of the algorithms have
    been used. This table refers to the Fortran version of our code; the C version is
    about 10\% slower.}
  \label{tab:speed}
\end{table}

On the other hand, our implementations of the iterative algorithms have a very simple
parameter set, they do not require any internal reorganization of data structures, avoid
function calls as far as possible, and contain some optimizations specific to the
$3 \times 3$ case. On the downside, they do not perform any pre-treatment of the
matrix such as rescaling it to avoid overflows and underflows. We believe that for
pathological matrices, LAPACK may be more accurate than our QL method.

It is interesting to observe that the iterative algorithms are slightly faster for
matrices with logarithmically distributed entries. The reason is that for many of these
matrices, the off-diagonal entries are already small initially, so fewer iterations
are required for the diagonalization. This is one of the reasons why QL is always faster
than Cuppen in the logarithmically distributed scenario, but may be slower in the linear
case. Another reason for this is the fact that logarithmically distributed matrices
are more likely to create situations in which the hybrid root finder, which is used
to solve Eq.~\eqref{eq:Cuppen-secular}, converges very slowly. This happens for example
if the analytically calculated starting values have large errors.

However, even in favorable situations, neither QL nor Cuppen can
compete with the analytical methods. These do not require any pre-treatment of
the matrix (like the transformation to tridiagonal form in the case of QL and Cuppen),
and though their implementation may look a bit lengthy due to the many special cases
that can arise, the number of floating point operations that are finally executed for
each matrix is very small.

If we compare the performance for real symmetric vs.\ complex hermitian matrices,
we find that purely real problems can be solved much more efficiently. This is
especially true for the Jacobi algorithm since real Jacobi transformations are much
cheaper than complex ones. For QL and Cuppen, the performance bonus is less dramatic
because large parts of these algorithms always operate on purely real data structures.
The Cardano and vector product algorithms contain complex arithmetic, but
their speed is also limited by the evaluation of trigonometric functions (Cardano)
and by several conditional branches (vector product), therefore they too do not
benefit as much as the Jacobi method.

\subsection{Numerical accuracy}

The excellent performance of the analytical and hybrid algorithms is relativized by the
results of our accuracy tests, which are shown in Table~\ref{tab:accuracy}. While for
linearly distributed matrix entries, all algorithms get close to the machine precision
of about $2 \cdot 10^{-16}$, the Cardano and vector product methods become unstable
for the logarithmically distributed case. In particular, the large average values of
$\Delta_3 > \mathcal{O}(10^{-3})$ show that the calculated eigenvalues and
eigenvectors often fail to fulfill their defining property
$\mat{A} \vec{v} = \lambda \vec{v}$. This problem is mitigated by the hybrid technique,
but even this approach is still far less accurate than the Jacobi, QL, and Cuppen
algorithms. For these, $\Delta_3$ is still of order $10^{-9}$
(QL \& Cuppen) resp.\  $10^{-10}$ (Jacobi). The fact that Jacobi is more accurate
than QL and Cuppen confirms our expectations from Sec.~\ref{sec:JacobiEff}.

Note that the values of $\Delta_1$ and $\Delta_2$ for the LAPACK QL algorithm
are zero in the case of complex matrices and extremely small for real matrices.
The reason is that LAPACK QL is the reference algorithm used in the definition of
these quantities. In the case of real matrices, $\Delta_1$ and $\Delta_2$ reveal the
tiny differences between the LAPACK \texttt{ZHEEV} and \texttt{DSYEV} routines.

It is interesting to observe that in the logarithmically distributed scenario,
$\Delta_1$ and $\Delta_2$ are systematically larger for real than for complex
matrices. This does not have a deeper reason but is simply due to the fact that
in the real case, there are fewer random degrees of freedom, so there is a higher
chance for ill-conditioned matrices to occur. The effect is not visible in
$\Delta_3$ because there it is compensated by the fact that this quantity receives
large contributions mainly when in the evaluation of $\mat{A} \tilde{\vec{v}}_i$ in
Eq.~\eqref{eq:Delta3}, a multiplication of a large matrix entry with a small and
thus relatively inaccurate vector component occurs. It follows from combinatorics
that this is more likely to happen if $\mat{A}$ and $\tilde{\vec{v}}$ are complex.

\begin{table*}[p]
  \centering
  \textbf{Linearly distributed real matrix entries}
  \vspace{0.2 cm}

  \begin{ruledtabular}
  \begin{tabular}{l|c|c||c|c||c|c}
                 & \multicolumn{2}{c||}{$\Delta_1$} & \multicolumn{2}{c||}{$\Delta_2$}
                     & \multicolumn{2}{c}{$\Delta_3$}            \\  \hline
     Algorithm   & Avg. & Max. & Avg. & Max. & Avg. & Max.       \\  \hline
     Jacobi          & $1.34 \cdot 10^{-15}$ & $3.52 \cdot 10^{ -9}$     
                     & $4.01 \cdot 10^{-16}$ & $1.32 \cdot 10^{-12}$     
                     & $2.01 \cdot 10^{-15}$ & $1.02 \cdot 10^{ -8}$ \\  
                                                                                  
     QL              & $1.89 \cdot 10^{-15}$ & $5.59 \cdot 10^{ -9}$     
                     & $4.09 \cdot 10^{-16}$ & $2.05 \cdot 10^{-12}$     
                     & $3.58 \cdot 10^{-15}$ & $1.24 \cdot 10^{ -8}$ \\  
                                                                                  
     Cuppen          & $1.95 \cdot 10^{-15}$ & $9.53 \cdot 10^{ -9}$     
                     & $6.83 \cdot 10^{-16}$ & $1.80 \cdot 10^{-12}$     
                     & $4.21 \cdot 10^{-15}$ & $1.45 \cdot 10^{ -8}$ \\  
                                                                                  
     GSL             & $1.29 \cdot 10^{-15}$ & $3.18 \cdot 10^{ -9}$     
                     & $3.56 \cdot 10^{-16}$ & $2.18 \cdot 10^{-12}$     
                     & $2.40 \cdot 10^{-15}$ & $5.02 \cdot 10^{ -9}$ \\  
                                                                                  
     LAPACK QL       & $5.80 \cdot 10^{-17}$ & $3.61 \cdot 10^{-11}$     
                     & $3.17 \cdot 10^{-17}$ & $6.10 \cdot 10^{-13}$     
                     & $2.69 \cdot 10^{-15}$ & $8.28 \cdot 10^{ -9}$ \\  
                                                                                  
     Analytical      & $1.87 \cdot 10^{-15}$ & $9.53 \cdot 10^{ -9}$     
                     & $6.19 \cdot 10^{-15}$ & $1.80 \cdot 10^{ -8}$     
                     & $1.36 \cdot 10^{-14}$ & $4.32 \cdot 10^{ -8}$ \\  

     Hybrid          & $1.87 \cdot 10^{-15}$ & $9.53 \cdot 10^{ -9}$     
                     & $4.91 \cdot 10^{-15}$ & $6.49 \cdot 10^{ -9}$     
                     & $1.16 \cdot 10^{-14}$ & $2.91 \cdot 10^{ -8}$ \\  
  \end{tabular}
  \end{ruledtabular}
  
  \vspace{0.5 cm}
  \textbf{Linearly distributed complex matrix entries}

  \vspace{0.2 cm}
  \begin{ruledtabular}
  \begin{tabular}{l|c|c||c|c||c|c}
                     & \multicolumn{2}{c||}{$\Delta_1$} & \multicolumn{2}{c||}{$\Delta_2$}
                         & \multicolumn{2}{c}{$\Delta_3$}            \\  \hline
     Algorithm       & Avg. & Max. & Avg. & Max. & Avg. & Max.       \\  \hline
     Jacobi          & $1.96 \cdot 10^{-15}$ & $7.66 \cdot 10^{ -9}$     
                     & $4.64 \cdot 10^{-16}$ & $1.13 \cdot 10^{-13}$     
                     & $1.42 \cdot 10^{-14}$ & $3.44 \cdot 10^{ -7}$ \\  
                                                                                  
     QL              & $2.08 \cdot 10^{-14}$ & $5.46 \cdot 10^{ -7}$     
                     & $4.83 \cdot 10^{-16}$ & $8.16 \cdot 10^{-14}$     
                     & $4.27 \cdot 10^{-14}$ & $1.14 \cdot 10^{ -6}$ \\  
                                                                                  
     Cuppen          & $4.37 \cdot 10^{-15}$ & $6.54 \cdot 10^{ -8}$     
                     & $6.60 \cdot 10^{-16}$ & $2.03 \cdot 10^{-13}$     
                     & $3.95 \cdot 10^{-14}$ & $1.03 \cdot 10^{ -6}$ \\  
                                                                                  
     GSL             & $8.01 \cdot 10^{-15}$ & $1.88 \cdot 10^{ -7}$     
                     & $4.56 \cdot 10^{-16}$ & $8.36 \cdot 10^{-14}$     
                     & $2.14 \cdot 10^{-14}$ & $5.26 \cdot 10^{ -7}$ \\  
                                                                                  
     LAPACK QL       & $      0.0          $ & $      0.0          $     
                     & $      0.0          $ & $      0.0          $     
                     & $2.41 \cdot 10^{-14}$ & $6.03 \cdot 10^{ -7}$ \\  
                                                                                  
     Analytical      & $4.19 \cdot 10^{-15}$ & $6.54 \cdot 10^{ -8}$     
                     & $5.66 \cdot 10^{-16}$ & $3.17 \cdot 10^{-11}$     
                     & $3.05 \cdot 10^{-14}$ & $7.95 \cdot 10^{ -7}$ \\  

     Hybrid          & $4.19 \cdot 10^{-15}$ & $6.54 \cdot 10^{ -8}$     
                     & $5.56 \cdot 10^{-16}$ & $3.17 \cdot 10^{-11}$     
                     & $3.03 \cdot 10^{-14}$ & $7.95 \cdot 10^{ -7}$ \\  
  \end{tabular}
  \end{ruledtabular}
  
  \vspace{0.5 cm}
  \textbf{Logarithmically distributed real matrix entries}

  \vspace{0.2 cm}
  \begin{ruledtabular}
  \begin{tabular}{l|c|c||c|c||c|c}
                 & \multicolumn{2}{c||}{$\Delta_1$} & \multicolumn{2}{c||}{$\Delta_2$}
                     & \multicolumn{2}{c}{$\Delta_3$}            \\  \hline
     Algorithm   & Avg. & Max. & Avg. & Max. & Avg. & Max.       \\  \hline
     Jacobi          & $2.96 \cdot 10^{-10}$ & $1.94 \cdot 10^{ -4}$     
                     & $3.05 \cdot 10^{-12}$ & $3.91 \cdot 10^{ -7}$     
                     & $8.16 \cdot 10^{-11}$ & $1.10 \cdot 10^{ -4}$ \\  
                                                                                  
     QL              & $4.88 \cdot 10^{-10}$ & $4.29 \cdot 10^{ -4}$     
                     & $2.59 \cdot 10^{-12}$ & $7.14 \cdot 10^{ -7}$     
                     & $1.03 \cdot 10^{ -9}$ & $1.18 \cdot 10^{ -3}$ \\  
                                                                                  
     Cuppen          & $4.28 \cdot 10^{-10}$ & $4.29 \cdot 10^{ -4}$     
                     & $3.58 \cdot 10^{-12}$ & $6.55 \cdot 10^{ -7}$     
                     & $8.90 \cdot 10^{-10}$ & $1.12 \cdot 10^{ -3}$ \\  
                                                                                  
     GSL             & $1.86 \cdot 10^{-10}$ & $1.62 \cdot 10^{ -4}$     
                     & $2.78 \cdot 10^{-12}$ & $4.01 \cdot 10^{ -7}$     
                     & $9.87 \cdot 10^{-10}$ & $2.04 \cdot 10^{ -3}$ \\  
                                                                                  
     LAPACK QL       & $8.36 \cdot 10^{-12}$ & $1.14 \cdot 10^{ -5}$     
                     & $1.28 \cdot 10^{-13}$ & $1.81 \cdot 10^{ -7}$     
                     & $1.11 \cdot 10^{ -9}$ & $1.18 \cdot 10^{ -3}$ \\  
                                                                                  
     Analytical      & $1.87 \cdot 10^{ -9}$ & $7.20 \cdot 10^{ -3}$     
                     & $1.80 \cdot 10^{ -7}$ & $1.36 \cdot 10^{ +0}$     
                     & $3.47 \cdot 10^{ -1}$ & $1.07 \cdot 10^{ +6}$ \\  

     Hybrid          & $1.40 \cdot 10^{ -9}$ & $1.16 \cdot 10^{ -3}$     
                     & $3.84 \cdot 10^{-11}$ & $2.03 \cdot 10^{ -4}$     
                     & $2.19 \cdot 10^{ -4}$ & $4.75 \cdot 10^{ +1}$ \\  
  \end{tabular}
  \end{ruledtabular}
  
  \vspace{0.5 cm}
  \textbf{Logarithmically distributed complex matrix entries}

  \vspace{0.2 cm}
  \begin{ruledtabular}
  \begin{tabular}{l|c|c||c|c||c|c}
                 & \multicolumn{2}{c||}{$\Delta_1$} & \multicolumn{2}{c||}{$\Delta_2$}
                     & \multicolumn{2}{c}{$\Delta_3$}            \\  \hline
     Algorithm   & Avg. & Max. & Avg. & Max. & Avg. & Max.       \\  \hline
     Jacobi          & $1.55 \cdot 10^{-10}$ & $1.64 \cdot 10^{ -4}$     
                     & $2.23 \cdot 10^{-13}$ & $7.43 \cdot 10^{ -8}$     
                     & $1.19 \cdot 10^{-10}$ & $8.24 \cdot 10^{ -5}$ \\  
                                                                                  
     QL              & $2.25 \cdot 10^{-10}$ & $6.84 \cdot 10^{ -4}$     
                     & $1.96 \cdot 10^{-13}$ & $1.17 \cdot 10^{ -7}$     
                     & $7.85 \cdot 10^{-10}$ & $5.93 \cdot 10^{ -4}$ \\  
                                                                                  
     Cuppen          & $2.03 \cdot 10^{-10}$ & $6.02 \cdot 10^{ -4}$     
                     & $2.71 \cdot 10^{-13}$ & $1.30 \cdot 10^{ -7}$     
                     & $7.59 \cdot 10^{-10}$ & $5.86 \cdot 10^{ -4}$ \\  
                                                                                  
     GSL             & $1.06 \cdot 10^{-10}$ & $8.69 \cdot 10^{ -5}$     
                     & $2.17 \cdot 10^{-13}$ & $1.30 \cdot 10^{ -7}$     
                     & $1.38 \cdot 10^{ -9}$ & $1.15 \cdot 10^{ -3}$ \\  
                                                                                  
     LAPACK QL       & $      0.0          $ & $      0.0          $     
                     & $      0.0          $ & $      0.0          $     
                     & $1.27 \cdot 10^{ -9}$ & $1.24 \cdot 10^{ -3}$ \\  
                                                                                  
     Analytical      & $1.16 \cdot 10^{ -9}$ & $7.10 \cdot 10^{ -4}$     
                     & $6.10 \cdot 10^{ -9}$ & $8.29 \cdot 10^{ -2}$     
                     & $2.88 \cdot 10^{ -3}$ & $2.84 \cdot 10^{ +3}$ \\  

     Hybrid          & $1.11 \cdot 10^{ -9}$ & $6.84 \cdot 10^{ -4}$     
                     & $8.55 \cdot 10^{-12}$ & $3.55 \cdot 10^{ -5}$     
                     & $1.15 \cdot 10^{ -4}$ & $8.81 \cdot 10^{ +1}$ \\  
  \end{tabular}
  \end{ruledtabular}
  \caption{Numerical accuracy of different algorithms for calculating the eigenvalues
    and eigenvectors of symmetric or hermitian $3 \times 3$ matrices. This table
    refers to the Fortran implementation, but we have checked that the values obtained
    with the C code are similar.}
  \label{tab:accuracy}
\end{table*}

\section{Conclusions}
\label{sec:conc}

In this article, we have studied the numerical three-dimensional eigenproblem for
symmetric and hermitian matrices. We have discussed the Jacobi, QL, and Cuppen
algorithms as well as an analytical method using Cardano's formula and
vector cross products. Our benchmarks reveal that standard packages
are very slow for small matrices. Optimized versions of the standard algorithms
are a lot faster while retaining similar numerical properties, but even their
speed is not competitive with that of the analytical methods. We have, however,
seen that the latter have limited numerical accuracy in extreme situations. Moreover,
they were not designed to avoid overflow and underflow conditions. To partly
circumvent these problems, we have devised a hybrid algorithm, which employs
analytical calculations as the standard branch, but falls back to QL if it estimates
the problem to be ill-conditioned.

Depending on what kind of problem is to be solved, we give the following recommendations:
\begin{itemize}
  \item \textbf{The hybrid algorithm} is recommended for problems where
    computational speed is more important than accuracy, and the matrices are not too
    ill-conditioned in the sense that their eigenvalues do not differ by more than
    a few orders of magnitude. For example, in the initial example of the neutrino
    oscillation Hamiltonian, where the physical uncertainties in the parameters are
    much larger than the numerical errors, the hybrid algorithm turns out to be the
    optimal choice~\cite{Huber:2007ji}.
  \item The \textbf{QL algorithm} is a good general purpose ``black box'' method
    since it is reasonably fast and --- except in some very special situations like
    the example given in Eq.~\eqref{eq:testmatrix1} --- also very accurate. If speed
    is not an issue, one can use standard implementations of QL like the LAPACK
    function \texttt{ZHEEV}. For better performance we recommend simpler implementations
    like our function \texttt{ZHEEVQ3} or the function \texttt{tqli} from
    Ref.~\cite{Press:NumRecip}, on which our routine is based.
  \item \textbf{Cuppen's Divide and Conquer method} can achieve an accuracy similar
    to that of the QL algorithm and may be slightly faster for complex problems
    if the input matrix is not already close to diagonal. The choice between Cuppen
    and QL will therefore depend on the details of the problem that is to be solved.
  \item If the highest possible accuracy is desired, \textbf{Jacobi's method} is the
    algorithm of choice. It is extremely accurate even for very pathological
    matrices, but it is significantly slower than the other algorithms, especially
    for complex problems.
  \item The \textbf{purely analytical method} is not recommended for practical
    applications because it is superseded by the hybrid algorithm. It is, however,
    of academic interest since it reveals both the strengths and the limitations
    of analytical eigensystem calculations.
\end{itemize}
Let us remark that in scenarios where the diganolization of a hermitian
$3 \times 3$ matrix is only part of a larger problem, it may be advantageous to
choose a slower but more accurate algorithm because this may improve the convergence
of the surrounding algorithm, thus speeding up the overall process. The final
choice of diagonalization algorithm will always depend strongly on the details
of the problem which is to be solved.

\section*{Acknowledgments}

I would like to thank M.~Lindner, P.~Huber, and the readers of the NA Digest
for useful discussion and comments. Furthermore I would like to acknowledge support
from the Studienstiftung des Deutschen Volkes.

\appendix
\section{Alternative derivations of Cardano's method}
\label{sec:cardano-alt}

In this appendix, we will discuss two alternative solution strategies for
the third degree polynomial equations~\eqref{eq:charpoly}, and show
that in the end, numerical considerations lead again to Eq.~\eqref{eq:cardano}.

\subsection{A trigonometric approach}

If we substitute $x = 2\cos\tfrac{\theta}{3}$ on the left hand side of
Eq.~\eqref{eq:charpoly-reduced} and use trigonometric transformations
to obtain
\begin{equation}
  2 \cos\theta = t,
\end{equation}
we can show that the solutions to Eq.~\eqref{eq:charpoly-reduced} can be
written as
\begin{equation}
  x = 2 \cos\tfrac{\theta}{3} = 2 \cos \Big( \tfrac{1}{3} \arccos\tfrac{t}{2} \Big).
  \label{eq:chebyshev}
\end{equation}
Our previous result $-2 \leq t \leq 2$ (see Sec.~\ref{sec:alg-noniter-values})
ensures that this is well-defined. If we replace the $\arccos$ by a numerically
more stable $\arctan$, we immediately recover Eq.~\eqref{eq:cardano}.

\subsection{Lagrange resolvents}

The second alternative to Cardano's derivation that we are going to consider employs the
concept of Lagrange resolvents. We start from the observation that the coefficients of
Eq.~\eqref{eq:charpoly} can be expressed in terms of the roots $\lambda_1$,
$\lambda_2$, and $\lambda_3$ of $P(\lambda)$, because we can write
$P(\lambda) = \prod_{i=1,2,3} (\lambda - \lambda_i)$. In particular, $c_0$, $c_1$, and
$c_2$ are the so-called \emph{elementary symmetric polynomials} in $\lambda_1$,
$\lambda_2$, and $\lambda_3$:
\begin{equation}
  \begin{aligned}
    -c_2 &= \sum \lambda_i,                  \\
     c_1 &= \sum_{i < j} \lambda_i\lambda_j, \\
    -c_0 &= \prod \lambda_i.
  \end{aligned}
\end{equation}

Next, we consider the \emph{Lagrange resolvents} of Eq.~\eqref{eq:charpoly}, which are
defined by
\begin{equation}
  \begin{aligned}
    r_1 &= \lambda_1 + \lambda_2 + \lambda_3,                                            \\
    r_2 &= \lambda_1 + e^{i\tfrac{2}{3}\pi} \lambda_2 + e^{-i\tfrac{2}{3}\pi} \lambda_3, \\
    r_3 &= \lambda_1 + e^{-i\tfrac{2}{3}\pi} \lambda_2 + e^{i\tfrac{2}{3}\pi} \lambda_3.
  \end{aligned}
\end{equation}
We observe, that $r_i^3$ is invariant under permutation of the $\lambda_i$,
and so, by the fundamental theorem of symmetric functions~\cite{vdWaerden:Algebra1},
can be expressed in terms of $c_0$, $c_1$, and $c_2$. Indeed, with the
definitions from Eq.~\eqref{eq:tpq}, we obtain
\begin{equation}
  \begin{aligned}
    r_1^3 &= -c_2^3,               \\
    r_2^3 &= q + \sqrt{q^2 - p^3}, \\
    r_3^3 &= q - \sqrt{q^2 - p^3}.
  \end{aligned}
\end{equation}
We can then recover $\lambda_1$, $\lambda_2$, and $\lambda_3$ according to
\begin{equation}
  \begin{aligned}
    \lambda_1 &= \tfrac{1}{3} (r_1 + r_2 + r_3),                                            \\
    \lambda_2 &= \tfrac{1}{3} (r_1 + e^{-i\tfrac{2}{3}\pi} r_2 + e^{i\tfrac{2}{3}\pi} r_3), \\
    \lambda_3 &= \tfrac{1}{3} (r_1 + e^{i\tfrac{2}{3}\pi} r_2 + e^{-i\tfrac{2}{3}\pi} r_3).
  \end{aligned}
\end{equation}

For a practical implementation of these formulas, one would like to avoid complex
arithmetic. This is possible because we have seen before that $\sqrt{q^2 - p^3}$
is always purely imaginary. This observation allows us to write
\begin{equation}
  \begin{aligned}
    r_2 &= \sqrt{p} e^{i \phi}, \\
    r_3 &= \sqrt{p} e^{-i \phi},
  \end{aligned}
\end{equation}
with $\phi = \tfrac{1}{3} \arctan \sqrt{p^3 - q^2}/q$ as before, and thus
\begin{equation}
  \begin{aligned}
    \lambda_1 &= \tfrac{1}{3} (-c_2 + 2\rho\cos\phi),                        \\
    \lambda_2 &= \tfrac{1}{3} (-c_2 - \rho\cos\phi - \sqrt{3} \rho\sin\phi), \\ 
    \lambda_3 &= \tfrac{1}{3} (-c_2 - \rho\cos\phi + \sqrt{3} \rho\sin\phi).
  \end{aligned}
\end{equation}
These expressions are equivalent to Eq.~\eqref{eq:cardano} after substituting
back Eqs.~\eqref{eq:tpq}, so the practical implementation of the
Lagrange method is again identical to the previous algorithms.

\section{Documentation of the C and Fortran code}
\label{sec:code}

Along with the publication of this article, we provide C and Fortran implementations
of the algorithms discussed here for download. They are intended to be used
for further numerical experiments or for the solution of actual scientific problems.

Our C code follows the C99 standard which provides the complex data type
\texttt{double~complex}.  In $\texttt{gcc}$, this requires the usage of the
compiler option $\texttt{-std=c99}$. The Fortran code is essentially Fortran 77,
except for the fact that not all variable and function names obey the 6-character
limitation.

Both versions of the code contain detailed comments, describing the structure
of the routines, the purpose of the different functions, and their arguments.
The C version also contains detailed information about local variables, which
was omitted in the Fortran version to keep the code compact.

Our nomenclature conventions for functions and subroutines may
seem a bit cryptical because we tried to keep as close as possible
to the LAPACK conventions: The first letter indicates the data type (``D'' for
double or ``Z'' for double complex), the second and third letters indicate the
matrix type (``SY'' for symmetric and ``HE'' for hermitian), while the remaining
characters specify the purpose of the function: ``EV'' means eigenvalues
and/or eigenvectors, ``J'' stands for Jacobi, ``Q'' for QL,
``D'' for Divide \& Conquer (Cuppen), ``V'' for vector product, and ``C'' for
Cardano. We also add the suffix ``3'' to indicate that our routines are
designed for $3 \times 3$ matrices.

In the following we will describe the interface of the individual routines.
We will discuss only those functions which are relevant to the complex case,
because their real counterparts are similar, with the data types
\texttt{COMPLEX*16} resp.\ \texttt{double complex} being replaced by
\texttt{DOUBLE PRECISION} resp.\ \texttt{double}.

Furthermore, we will only discuss the Fortran code here because the
corresponding C functions have identical names and arguments. For example the
Fortran subroutine
\addtolength{\leftmargini}{-0.5 cm}
\begin{quote}
  \hangindent=0.3cm \hangafter=1 \noindent
     \texttt{SUBROUTINE ZHEEVJ3(A, Q, W) \\
               COMPLEX*16 A(3, 3) \\
               COMPLEX*16 Q(3, 3) \\
               DOUBLE PRECISION W(3)}
\end{quote}
corresponds to the C function
\begin{quote}
  \hangindent=0.3cm \hangafter=1 \noindent
    \texttt{int zheevj3(double~complex~A[3][3], double~complex~Q[3][3],
              double~w[3])}.
\end{quote}
\addtolength{\leftmargini}{+0.5 cm}

\subsection{Main driver function}

\begin{refentry}{SUBROUTINE ZHEEVJ3(A, Q, W) \\
                   COMPLEX*16 A(3, 3) \\
                   COMPLEX*16 Q(3, 3) \\
                   DOUBLE PRECISION W(3) }
  
  This routine uses Jacobi's method (see Sec.~\ref{sec:jacobi}) to find the eigenvalues
  and normalized eigenvectors of a hermitian $3 \times 3$ matrix $\texttt{A}$.
  The eigenvalues are stored in \texttt{W}, while the eigenvectors are returned
  in the columns of \texttt{Q}.

  The upper triangular part of \texttt{A} is destroyed during the calculation,
  the diagonal elements are read but not destroyed, and the lower triangular elements
  are not referenced at all.
\end{refentry}

\begin{refentry}{SUBROUTINE ZHEEVQ3(A, Q, W) \\
                   COMPLEX*16 A(3, 3) \\
                   COMPLEX*16 Q(3, 3) \\
                   DOUBLE PRECISION W(3) }

  This is our implementation of the QL algorithm from Sec.~\ref{sec:ql}. It finds
  the eigenvalues and normalized eigenvectors of a hermitian $3 \times 3$ matrix
  $\texttt{A}$ and stores them in \texttt{W} and in the columns of \texttt{Q}.

  The function accesses only the diagonal and upper triangular parts of \texttt{A}.
  The access is read-only.
\end{refentry}

\begin{refentry}{SUBROUTINE ZHEEVD3(A, Q, W) \\
                   COMPLEX*16 A(3, 3) \\
                   COMPLEX*16 Q(3, 3) \\
                   DOUBLE PRECISION W(3) }

  This is Cuppen's Divide and Conquer algorithm, optimized for 3-dimensional problems
  (see Sec.~\ref{sec:cuppen}). The function assumes \texttt{A} to be a hermitian
  $3 \times 3$ matrix, and calculates its eigenvalues $\texttt{W}_i$, as
  well as its normalized eigenvectors. The latter are returned in the columns of
  \texttt{Q}.

  The function accesses only the diagonal and upper triangular parts of \texttt{A}.
  The access is read-only.
\end{refentry}

\begin{refentry}{SUBROUTINE ZHEEVC3(A, W) \\
                   COMPLEX*16 A(3, 3) \\
                   DOUBLE PRECISION W(3) }

  This routine calculates the eigenvalues $\texttt{W}_i$ of a hermitian $3 \times 3$
  matrix $\texttt{A}$ using Cardano's analytical algorithm
  (see Sec.~\ref{sec:alg-noniter-values}).
  Only the diagonal and upper triangular parts of $\texttt{A}$ are accessed,
  and the access is read-only.
\end{refentry}

\begin{refentry}{SUBROUTINE ZHEEVV3(A, Q, W) \\
                   COMPLEX*16 A(3, 3) \\
                   COMPLEX*16 Q(3, 3) \\
                   DOUBLE PRECISION W(3) }

  This function first calls \texttt{ZHEEVC3} to find the eigenvalues of the
  hermitian $3 \times 3$ matrix $\texttt{A}$, and then uses vector cross products
  to analytically calculate the normalized eigenvectors
  (see Sec.~\ref{sec:alg-noniter-vectors}). The eigenvalues are stored in \texttt{W},
  the normalized eigenvectors in the columns of \texttt{Q}.
  
  Only the diagonal and upper triangular parts of \texttt{A} need to contain meaningful
  values, but all of \texttt{A} may be used as temporary storage and might hence be
  destroyed.
\end{refentry}

\begin{refentry}{SUBROUTINE ZHEEVH3(A, Q, W) \\
                   COMPLEX*16 A(3, 3) \\
                   COMPLEX*16 Q(3, 3) \\
                   DOUBLE PRECISION W(3) }

  This is the hybrid algorithm from Sec.~\ref{sec:alg-noniter-hybrid}. Its default
  behavior is identical to that of \texttt{ZHEEVV3}, but under certain circumstances,
  it falls back to calling \texttt{ZHEEVQ3}. As for the other routines,
  $\texttt{A}$ has to be a hermitian $3 \times 3$ matrix, and the eigenvalues and
  eigenvectors are stored in \texttt{W} resp.\ in the columns of \texttt{Q}.
  
  Only the diagonal and upper triangular parts of \texttt{A} need to contain meaningful
  values, and access to \texttt{A} is read-only.
\end{refentry}

\subsection{Helper functions}

\begin{refentry}{SUBROUTINE DSYEV2(A, B, C, RT1, RT2, CS, SN) \\
                   DOUBLE PRECISION A, B, C \\
                   DOUBLE PRECISION RT1, RT2, CS, SN }

  This subroutine calculates the eigenvalues and eigenvectors of a real symmetric
  $2 \times 2$ matrix
  \begin{equation}
    \begin{pmatrix}
      \texttt{A}  &  \texttt{B}  \\
      \texttt{B}  &  \texttt{C}
    \end{pmatrix}.
  \end{equation}
  The result satisfies
  \begin{equation}
    \begin{pmatrix}
      \texttt{RT1} &   0         \\
            0      & \texttt{RT2}
    \end{pmatrix} =
    \begin{pmatrix}
      \texttt{CS}  & \texttt{SN}   \\
      \texttt{-SN} & \texttt{CS}
    \end{pmatrix}
    \begin{pmatrix}
      \texttt{A}  &  \texttt{B}  \\
      \texttt{B}  &  \texttt{C}
    \end{pmatrix}
    \begin{pmatrix}
      \texttt{CS}  & \texttt{-SN}  \\
      \texttt{SN}  & \texttt{CS}
    \end{pmatrix}
  \end{equation}
  and $\texttt{RT1} \geq \texttt{RT2}$. Note that this convention is different
  from the convention used in the corresponding LAPACK function \texttt{DLAEV2},
  where $|\texttt{RT1}| \geq |\texttt{RT2}|$. We use a different convention
  here because it helps to avoid several conditional branches in \texttt{ZHEEVD3}
  and \texttt{DSYEVD3}.
\end{refentry}

\begin{refentry}{SUBROUTINE ZHETRD3(A, Q, D, E) \\
                   COMPLEX*16 A(3, 3) \\
                   COMPLEX*16 Q(3, 3) \\
                   DOUBLE PRECISION D(3) \\
                   DOUBLE PRECISION E(2) }

  This routine reduces a hermitian matrix $\texttt{A}$ to real
  tridiagonal form by applying a Householder transformation $\texttt{Q}$
  according to Sec.~\ref{sec:ql}:
  \begin{equation}
    \texttt{A} = \texttt{Q} \cdot \begin{pmatrix}
                                    \texttt{D}_1 & \texttt{E}_1 &              \\
                                    \texttt{E}_1 & \texttt{D}_2 & \texttt{E}_2 \\
                                                 & \texttt{E}_2 & \texttt{D}_3
                                  \end{pmatrix} \cdot \texttt{Q}^T.
  \end{equation}
  The function accesses only the diagonal and upper triangular parts of \texttt{A}.
  The access is read-only.
\end{refentry}



\end{document}